# Chapter 13

## Quantum Capacitance of Graphene Sheets and Nanoribbons

**George S. Kliros**

Department of Aeronautical Sciences, Division of Electronics, Electric Power and Telecommunications, Hellenic Air-Force Academy, Dekeleia Air-Force Base GR-1010, Attica, Greece.

**Abstract**

In this chapter, semi-analytical models for the calculation of the quantum capacitance of both monolayer and bilayer graphene and its nanoribbons, are presented. Since electron-hole puddles are experimental facts in all graphene samples, they have been incorporated in our calculations. The temperature dependence of the quantum capacitance around the charge neutrality point is also investigated and the obtained results are in agreement with many features recently observed in quantum capacitance measurements on both monolayer and bilayer graphene devices. Furthermore, the impact of finite-size and edge effects on the quantum capacitance of graphene nanoribbons is studied taking into account both the edge bond relaxation and third-nearest-neighbour interaction in the band structure of GNRs.

### 13.1 Introduction

One of the main characteristics of field effect transistors (FETs) is the capacitance formed between the channel and the gate. The gate capacitance is important for understanding fundamental electronic properties of the channel material such as the density of states (DOS) as well as the device performance including the I-V characteristics and the device operation frequency. Low-dimensional systems, having a small DOS, are not able to accumulate enough charge to completely screen the external field. In order to describe the effect of the electric field penetration through a two-dimensional electron gas (2DEG), Luryi introduced the concept of quantum capacitance $C_Q$ which describes the movement of the conduction band as function of applied gate voltage (Luryi et al. 1988, John et al. 2004). Consequently, the gate capacitance is the series connection of the insulator capacitance $C_{ins}$ and the quantum capacitance. In the 2DEG in semiconductor heterostructures, experiments have been accomplished mapping the DOS directly via the quantum capacitance and are nowadays standard characterization tools for these structures (Smith et al. 1985). The quantum capacitance in semiconductor superlattices has been studied during the last years revealing a number of interesting properties including the oscillating behaviour of the DOS in the presence of a magnetic field (Kliros and Divari 2007a) and signatures of spin-orbit interaction (Kliros and Divari 2007b, Kliros 2009). However, in such structures the insulator capacitance is much smaller than the corresponding quantum capacitance and as a result, the quantum capacitance is usually a small contribution that is difficult to discern experimentally. On the other hand, the energy spectrum of carbon nanotubes was experimentally observed in measurements of the quantum capacitance (Ilani et al. 2006).

Graphene has an atomically thin body so that its quantum capacitance can dominate the device's electrostatics. Moreover, its DOS is a strong function of Fermi energy and therefore, quantum capacitance can be changed by applying a gate voltage. Recently, the quantum capacitance of a graphene sheet has been measured by utilizing an ionic liquid as the gate insulator (Xia et al. 2009) or an ultrathin gate dielectric (Ponomarenko et al. 2010, Xu et al. 2011) and satisfactory agreement between theory and experimental data has been found at large bias far from the Dirac point. However, deviation from theory at small bias has been observed in all measurements. This deviation has been attributed to density fluctuations around the Dirac point due to electron-hole puddles induced by charged impurities (Martin et al. 2008). The measured quantum capacitance of bilayer graphene has been shown similar characteristics to that of monolayer graphene but, near the Dirac point, a finite capacitance value has been found (Xia et al. 2009). A unique feature of both monolayer and bilayer graphene is that, the density of carriers can be tuned continuously by an external gate from electron-like carriers at positive doping to hole-like at negative doping. An important difference between monolayer and bilayer graphene is the band structure near the Dirac point. Monolayer graphene has a conical band structure and a density of states that vanishes linearly at the Dirac point. Bilayer graphene has a hyperbolic band structure and a density of states rising linearly with increasing energy from a finite value at zero energy. Moreover, bilayer graphene has attracted great interest due to the fact that a sizable band gap can be opened by chemical doping or by applying an external electric field normal to the graphene plane (Zhang et al. 2009). On the other hand, if graphene is patterned in nanoribbons, a sizable bandgap opens due to the quantum confinement effect in its transverse direction (Son et al. 2010). Recent theoretical as well as experimental studies have shown that a narrow graphene nanoribbon (GNR) will become a semiconductor regardless of its chirality, since the width confinement and edge effect would open a large bandgap in its band structure. Due to this property, field effect transistors with GNR channels showing complete switch off and improved on-off current ratios, can be utilized as building blocks for future digital circuits (Kliros 2013). Moreover, the effects of uniaxial tensile strain on the ultimate performance of a dual-gated graphene nanoribbon field-effect transistor have been recently studied (Kliros 2014).
In graphene, the DOS is a function of Fermi energy and as a consequence, the capacitance of a metal-oxide-graphene capacitor increases with increasing carrier concentration. Utilizing this property in combination with the high mobility of graphene, we can design high quality factor variable capacitors (Koester 2011). The quantum capacitance effect in graphene sheets and their nanoribbons has been recently utilized to realize super capacitors, varactors for wireless sensing applications and various types of high performance sensors (Rumyantsev et al. 2012). The advantages of graphene quantum capacitance wireless sensors compared to alternative passive sensing devices include high noise immunity, improved size scalability, fast response and capability for sensing a wide range of species. Moreover, there has been a recent advancement in the applicability of graphene nanoribbons as interconnects in VLSI circuits (Li et al. 2009) as well as all-electronic ultra-high frequency oscillators and filters which are based on the distributed quantum capacitance, inductance and resistance (Begliarbekov et al. 2011). The impedance offered by these parameters is directly related to the graphene bandstructure and determines the switching performance of the device. The most important research works on the quantum capacitance of graphene sheets and its nanoribbons, are summarized in Table 1.1.

*Table 1.1: Research works on Quantum Capacitance of Monolayer Graphene Sheets and Nanoribbons*

| *Research work on* | *Authors* | *Year* |
|---|---|---|
| Carrier statistics and quantum capacitance of graphene sheets and ribbons. | Fang T. et al | 2007 |
| Gate electrostatics and quantum capacitance of graphene nanoribbons. | Guo J. et al | 2007 |
| Quantum capacitance in high performance graphene field-effect transistor devices. | Chen, Z. H. and Appenzeller, J. | 2008 |
| Measurement of the quantum capacitance of graphene. | Xia J. et al | 2009 |
| Quantum capacitance in graphene by scanning probe microscopy. | Giannazzo F. et al | 2009 |
| Capacitance of graphene nanoribbons. | Shylau A. et al. | 2009 |
| Density of states and zero Landau level probed through capacitance of graphene. | Ponomarenko L.A. et al | 2010 |
| Quantum Capacitance and Density of States of Graphene. | Droscher S. et al | 2010 |
| Quantum Capacitance Limited Vertical Scaling of Graphene Field-Effect Transistor. | Xu H. et al. | 2011 |
| Measurements and microscopic model of quantum capacitance in graphene. | Xu H. et al. | 2011 |
| Influence of density inhomogeneity on the quantum capacitance of graphene. | Kliros G.S. | 2012 |
| Interaction phenomena in graphene seen through quantum capacitance. | Yu G.L. et al. | 2013 |
| Electron-electron interactions in monolayer graphene quantum capacitors. | Chen X. et al. | 2013 |
| Capacitance of strongly disordered graphene. | Li W. et al | 2013 |
| Negative quantum capacitance in graphene nanoribbons with lateral gates. | Reiter R. et al | 2014 |
| Quantum capacitance in graphene for device scaling. | Lee J. et al. | 2014 |

In order to provide physical insight into the quantum capacitance of graphene, it is important to develop intuitive analytical models capturing the essential physics of the device at hand. In this chapter, we present semi-analytical models for the quantum capacitance of both monolayer (MLG) and bilayer graphene (BLG) devices. The quantum capacitance of graphene nanoribbons (GNRs) is also considered. The calculations incorporate the presence of electron-hole puddles induced by local potential fluctuations as well as finite temperature effects. The temperature dependence of the quantum capacitance around the charge neutrality point is also investigated. The obtained results are in agreement with many features recently observed in quantum capacitance measurements on both monolayer and bilayer graphene devices. Moreover, calculations of the quantum capacitance of GNRs are presented and the effect of density inhomogeneity due to the presence of electron-hole puddles, is explored. Finally, the impact of chirality and edge effects on the capacitance of the GNRs are studied by incorporating both edge bond relaxation and third-nearest-neighbour (3NN) interaction in the band structure of GNRs.

**13.2 Quantum Capacitance of Monolayer Graphene**

Monolayer graphene is a zero-gap semiconductor because its conduction and valence π-electron bands touch each other only at two isolated points in its two-dimensional Brillouin zone. The dispersion relation of these bands in the vicinity of these points is given by

$$E_s(k_x, k_y) = \pm \hbar v_F \sqrt{k_x^2 + k_y^2} \qquad (13.1)$$

where (+) stands for the conduction band (CB) and (-) for the valence band (VB), $v_F = \sqrt{3}\gamma_0 a / 2$, is the Fermi velocity with intralayer coupling $\gamma_0$=3.16 eV and $(k_x,k_y)$ is the wave vector of carriers in the two-dimensional plane of the graphene sheet. The point $(k_x,k_y) = (0,0)$, referred as the "Dirac point", is a convenient choice for the energy reference, i.e., $E(k_x=0, k_x=0)=0$ eV. Given the energy spectrum, the DOS is defined as

$$D(E) = \sum_k \delta(E - E(k)) \qquad (13.2)$$

By using the MLG dispersion and changing the summation into an integral, Eq. (41.2) reduces to

$$D(E) = \frac{g_s g_v}{2\pi(\hbar v_F)^2} |E| \qquad (13.3)$$

where the spin $g_s$ and valley $g_v$ degeneracies are taken into account.
The quantum capacitance is defined as the derivative of the total net charge of the graphene sheet with respect to applied electrostatic potential defined as $V_{ch} = E_F/e$, where $E_F$ denotes the Fermi energy and e is the proton charge. The total net charge is proportional to the weighted average of the DOS at the Fermi level $E_F$. When the DOS as a function of energy is known, the quantum capacitance $C_Q(V_{ch})$ of the channel at finite temperature can be calculated as (Knoch 2008).

$$C_Q(V_{ch}) = e^2 \int_{-\infty}^{+\infty} D(E)\left(-\frac{\partial f(E-E_F)}{\partial E}\right) dE \qquad (13.4)$$

where $f(E)$ is the Fermi-Dirac distribution. Inserting the expression (41.3) into Eq. (41.4) and evaluating the derivative of the Fermi-Dirac function, we obtain

$$C_Q(V_{ch}) = \frac{g_s g_v e^2}{2\pi \hbar^2 v_F^2} \left\{ \int_0^{+\infty} \frac{E e^{(-E-E_F)/k_B T}}{\left(1+e^{(-E-E_F)/k_B T}\right)^2} dE + \int_0^{+\infty} \frac{E e^{(E-E_F)/k_B T}}{\left(1+e^{(E-E_F)/k_B T}\right)^2} dE \right\} \qquad (13.5)$$

Integrating the above expression by parts, we obtain after some algebra

$$C_Q(V_{ch}) = \frac{g_s g_v e^2 k_B T}{2\pi \hbar^2 v_F^2} \Gamma(2)\left[F_0(-\eta_F) + F_0(\eta_F)\right] \qquad (13.6)$$

where $\Gamma(x)$ is the Gamma function, $\eta_F = E_F/(k_B T)$ and $F_0(\eta) = \ln(1+e^\eta)$.

In the low temperature limit, Eq. (13.6) reduces to a simple expression in agreement with the literature (Fang et al. 2007)

$$C_Q(V_{ch}) = \frac{2e^2 k_B T}{\pi \hbar^2 v_F^2} \ln\left[2 + 2\cosh\left(\frac{eV_{ch}}{k_B T}\right)\right] \qquad (13.7)$$

where the values $g_s=2$ and $g_v=2$ have been adopted.

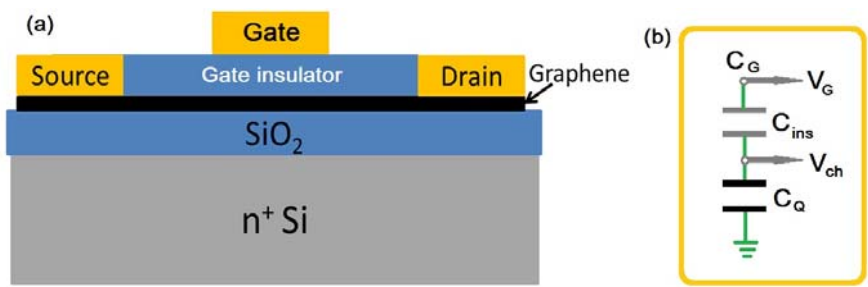

**Figure 13.1.** The geometry of a top-gated graphene device (a) and the equivalent circuit for quantum capacitance extraction (b).

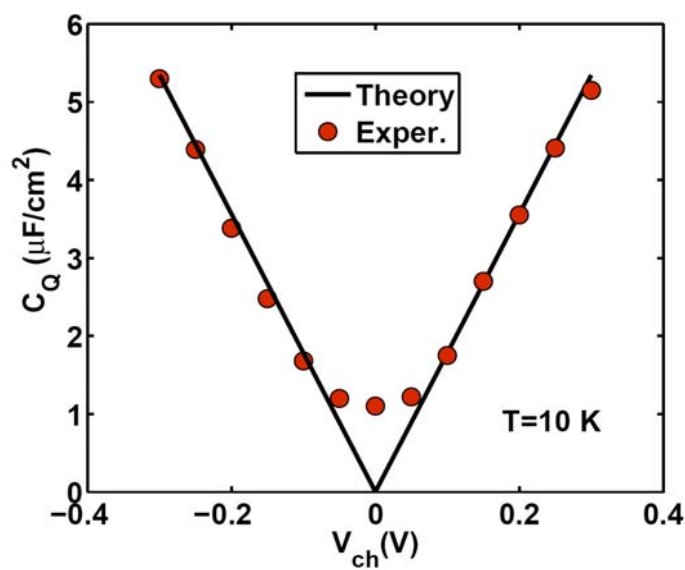

**Figure 13.2.** Theoretical quantum capacitance of pristine graphene at T=10 K along with the experimental data reported by (Ponomarenko et. al. 2010).

The first experimental studies on the quantum capacitance of MLG (Xia et al. 2009) have been performed using an electrochemical gate where the top surface of graphene sheet is exposed to an ionic liquid electrolyte. This electrochemical configuration makes the quantum capacitance a dominant component of the measured capacitance. However, in recent studies, the quantum capacitance has been measured using top-gated devices with a graphene channel and high-k thin gate insulators (e.g. $AlO_x$, $Y_2O_3$) as it is depicted in Fig. 13.1. The quantum capacitance $C_Q$ is extracted by the measured total gate capacitance $C_g$ that is a series combination of gate insulator capacitance $C_{ins}$ and $C_Q$, while the contribution due to paracitic capacitances is negligible and can be neglected, i.e. $C_Q = (C_g^{-1} - C_{ins}^{-1})^{-1}$. Fig. 13.2 shows the quantum capacitance as function of channel potential $V_{ch}$ along with the experimental data reported by (Ponomarenko et. al. 2010). As it is seen, the quantum capacitance $C_Q$ increases linearly with $V_{ch}$ in the region away from the neutrality point (Dirac point) with a slope of 18 µF cm$^{-2}$ V in agreement with the theoretical result. Moreover, the curve $C_Q(V_{ch})$ shows symmetric features for hole and electron branches around the Dirac point revealing the symmetry of the band structure of monolayer graphene. However, a large diviation from theory is seen near the Dirac point, that is, the slope becomes smaller and the measured minimum is much greater than the calculated minimum value $C_{Q,min} = 2e^2 k_B T \ln 4 /(\pi \hbar^2 v_F^2)$, i.e. ~0.8 µF/cm$^2$. This deviation suggest that more carriers exist around the Dirac point than the theoretically predicted and this fact can be explained by the density inhomogeneity due to the formation of electron-hole puddles (Droscher et al. 2010) induced by charged impurities around the graphene channel. It should be noted that, as the temperature increases, thermal excitation also begins to play important role.

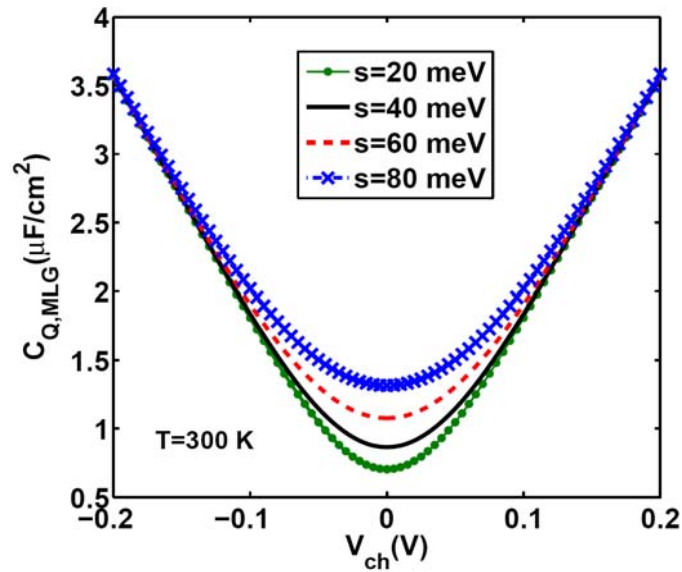

**Figure 13.3.** Calculated quantum capacitance for different potential fluctuation strengths at room temperature T=300 K.

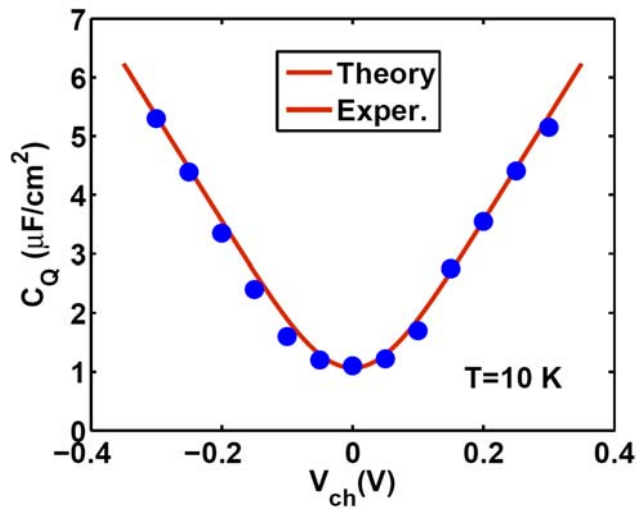

**Figure 13.4**. Fitting of the calculated quantum capacitance of MLG including electron-hole puddles to the measured one from (Ponomarenko et al. 2010).

Several studies have reported and experimentally confirmed that, unitended charged impurities present in the substrate-graphene interface, lead to the presence of inhomogeneous carrier density (i.e., electron-hole puddles) in the system (Martin et al. 2008, Li et al. 2011, Das Sarma et al. 2011). A rigorous way to incorporate the effects of potemtial fluctuations induced by the electron-hole puddle formation, is to introduce a local density of states (LDOS) from which a 'local' carrier density can be calculated. However, the simplest way is to assume a Gaussian distribution function for the density fluctuations associated with the puddles (Hwang and Das Sarma 2010). Then, the DOS of the graphene sheet in the presence of electron-hole puddles can be written as a convolution of the DOS of pristine graphene $D(\varepsilon)$ and the distribution function (Kliros 2010a, Kliros 2010b)

$$g(E) = \frac{1}{\sqrt{2\pi}s} \int_{-\infty}^{+\infty} \exp\left(-\frac{(\varepsilon-E)^2}{2s^2}\right) D(\varepsilon)\, d\varepsilon \qquad (13.8)$$

where $D(\varepsilon)$ is given by Eq. (13.3) and the parameter $\sigma$ denotes the standard deviation of the potential distribution representing the strength of the potential fluctuations and is the only the phenomenological parameter of our model. It is worth noting that the assumption of a Gaussian potential distribution is a good quantitative approximation to the actual numerically calculated inhomogeneous landscape of puddles in graphene. The potential fluctuation strength $s$ has been found to be in the range of 10 - 80 meV in typical graphene samples as extracted by fitting a recent microscopic self-consistent theory to existing experimental transport data (Li et al. 2011). These values of potential fluctuations are also consistent with direct numerical calculations of graphene electronic structure in the presence of quenched charged impurities (Das Sarma et al. 2010). After performing the integration in Eq. (13.8), we obtain

$$g_{MLG}(E) = \frac{2}{\pi(\hbar v_F)^2}\left[\sqrt{\frac{2}{\pi}}\, s\, exp\left(-\frac{E^2}{2s^2}\right) + E\, erf\left(\frac{E}{s\sqrt{2}}\right)\right] \qquad (13.9)$$

where $erf(x)$ is the Gaussian error function. Consequently, the quantum capacitance of MLG can be calculated as

$$C_{Q,MLG}(V_{ch}) = e^2 \int_{-\infty}^{+\infty} g_{MLG}(E)\left(-\frac{\partial f(E-E_F)}{\partial E}\right) dE \qquad (13.10)$$

In the low-temperature limit, the quantum capacitance can be approximated by

$$C_{Q,MLG}(T\approx 0) = e^2 g_{MLG}(E_F) = \frac{2e^2}{\pi(\hbar v_F)^2}\left[\sqrt{\frac{2}{\pi}}\, s\, exp\left(-\frac{E_F^2}{2s^2}\right) + E_F\, erf\left(\frac{E_F}{\sqrt{2}\,s}\right)\right] \qquad (13.11)$$

At the Dirac point ($V_{ch} = E_F/e = 0$) we get the minimum quantum capacitance

$$C_{Q,min}^{MLG}(T\approx 0, E_F \approx 0) = \frac{e^2 s}{(\hbar v_F)^2}\left(\frac{2}{\pi}\right)^{3/2} \qquad (13.12)$$

It is worth noting that, at the Dirac point electrons and holes are equally occupied. However, as the Fermi energy increases, more electrons occupy progressively larger portion of space and, for $E_F \gg s$, nearly all space is populated by the electrons so that the quantum capacitance of the system approaches that of the homogeneous material. Figure 13.3 shows our numerical results for the quantum capacitance for different potential fluctuation strengths at room temperature T=300 K. The numerical results from Eq. (13.10) along with the measurements of (Ponomarenko et al. 2010) are shown in Fig. 13.4. The standard deviation of potential fluctuations, $s$, providing the best fit to experimental data is found to be $s$=75 meV which is consistent with previous theoretical predictions (Droscher et al. 2010, Kliros 2010, Li et al. 2011). Fimally, the temperature dependence of the quantum capacitance, adopting $s$=75 meV, is explored in Fig. 13.5

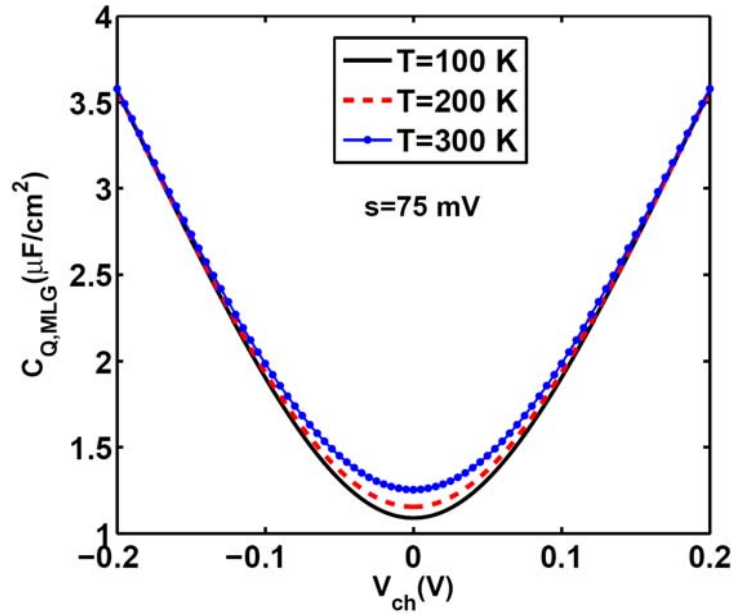

**Figure 13.5.** Temperature dependence of the quantum capacitance of MLG including electron-hole puddles. A value of $s$=75 meV is used in the calculations.

### 13.2 Quantum Capacitance of Bilayer Graphene

Graphene can be stacked in different forms but the only stable configuration is the Bernal stacking structure where a pair of honeycomb lattices, which include $A_1$ and $B_1$ carbon atoms on layer 1 and $A_2$ and $B_2$ on layer 2, are arranged so that $A_2$ atoms are located directly below $B_1$ atoms, as shown in Fig. 41.6a. The lattice constant within a layer is given by $a$=0.246 nm, the interlayer coupling $\gamma_1$=0.39 eV and the layer spacing by $d$=0.334 nm. The tight binding model for pristine BLG leads to a hyperbolic energy dispersion. In the absence of disorder, the band structure of pristine bilayer graphene can be written (McCann et al. 2006).

$$E_{\lambda,\mu}(k) = \lambda\left(\mu\frac{\gamma_1}{2} + \sqrt{\frac{\gamma_1^2}{4} + (\hbar v_F k)^2}\right) \qquad (13.13)$$

where $k = \sqrt{k_x^2 + k_y^2}$, $\lambda = \pm 1$, $\mu = \pm 1$ and $\gamma_1 = 0.39$ eV is the interlayer coupling. The index $\mu = (-)$ gives a pair of bands closer to zero energies, and $\mu = (+)$ another pair repelled away by approximately $\pm\gamma_1$. In each pair, the parameter $\lambda = (+1)$ and $(-1)$ represent the electron (CB) and hole (VB) branches, respectively.

It can be seen (Fig. 13.6b) that this energy dispersion interpolates between an approximately linear dispersion at large momentum to a quadratic one at small momentum, where the effective mass is $m^* = \gamma_1 / 2v_F^2$, that is, proportional to interlayer coupling. This crossover occurs at momentum $\hbar k = \gamma_1 / 2v_F$. An important difference between MLG and BLG is the band structure near the Dirac point. MLG has a conical band structure and a DOS that vanishes linearly at the Dirac point while BLG has a constant DOS at the Dirac point. However, recent experimental data have revealed a hyperbolic and asymmetric band structure without a constant DOS as expected for a quadratic dispersion (Henriksen and Eisenstein, 2010). Figs. 13.6 show the crystal lattice and the low energy bands of BLG whereas Fig. 13.7 illustrates a top-gated bilayer graphene device that can be used to perform BLG quantum capacitance studies. Recent investigations related to the quantum capacitance of BLG are summarized in Table 2.1.

*Table 2.1: Recent works on Quantum Capacitance of Bilayer Graphene*

| *Research work on* | *Authors* | *Year* |
| --- | --- | --- |
| Measurement of the electronic compressibility of bilayer graphene. | Henriksen, E. and Eisenstein, J. | 2010 |
| Model for the Quantum Capacitance of Bilayer Graphene Devices. | Kliros, G. S. | 2010 |
| Capacitance of graphene bilayer as a probe of layer-specific properties. | Young, A. F. and Levitov L. S. | 2011 |
| Electronic compressibility of layer-polarized bilayer graphene. | Young, A. F. et al | 2012 |
| Quantum capacitance in bilayer graphene nanoribbon. | Bhattacharya S. and Mahapatra S. | 2012 |
| Classic and Quantum Capacitances in Bernal Bilayer. | Sadeghi H. et al. | 2013 |
| Capacitance Variation of Electrolyte-Gated Bilayer Graphene Based Transistors. | Karimi Hediyeh et al. | 2013 |
| Tunable quantum capacitance and magnetic oscillation in bilayer graphene device. | Liu C. and Zhu J-L. | 2013 |
| Quantum Capacitance Measurement of Bilayer Graphene. | Nagashio K. et al. | 2014 |

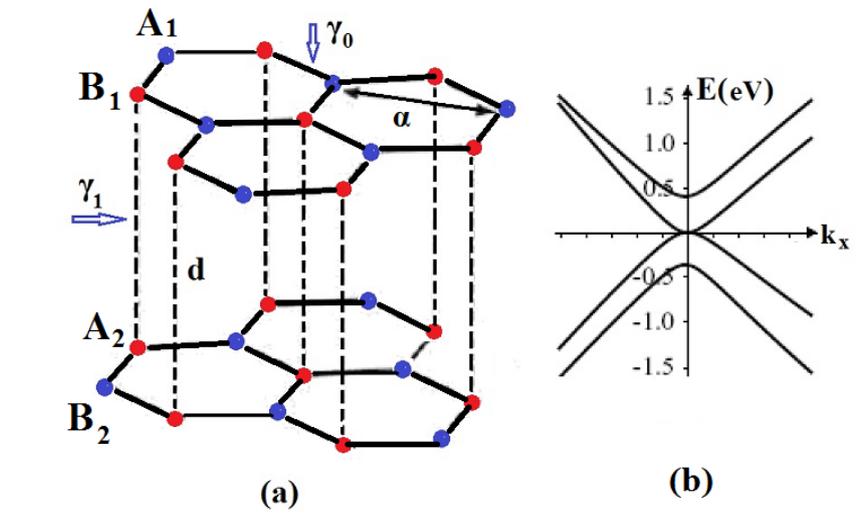

**Figure 13.6.** Crystal lattice and the low energy bands of bilayer graphene.

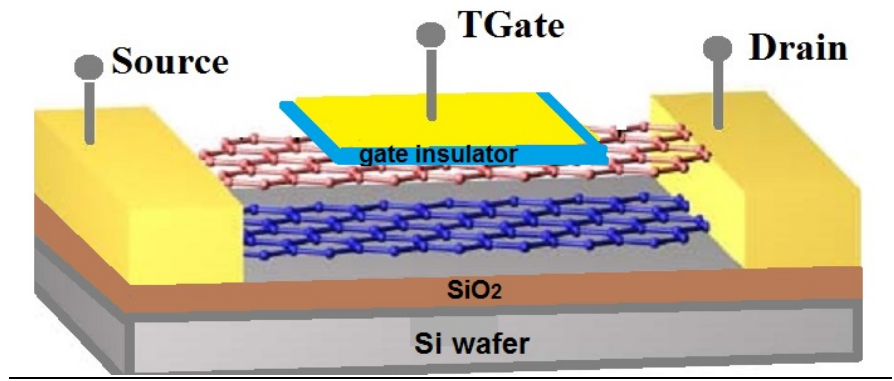

**Figure 13.7.** Schematics of a top-gated bilayer graphene device that can be used to perform quantum capacitance measurements. The BLG/thin top-gate dielectric stack lies on a $SiO_2$/Si substrate.

In the energy range $|E| \leq \gamma_1$, the DOS of the pure and perfect bilayer graphene can be well approximated by a linear relation as a function of energy (Barbier et al. 2009).

$$g_{BLG}(E) = \frac{g_s g_v}{2\pi(\hbar v_F)^2}\left(|E| + \frac{\gamma_1}{2}\right) \qquad (13.14)$$

where $g_s$, $g_v$ is the spin and valley degeneracy respectively. The above equation is accurate enough for low to moderate doping levels such that the Fermi level is less than 1 eV and is only incurs a relative error of up to a few percents when is between 1 eV and 2 eV. Near the Dirac point, the DOS can be written

$$g_{BLG}(E) = \frac{g_s g_v}{2\pi(\hbar v_F)^2} \frac{\gamma_1}{2} = g_s g_v \frac{m^*}{2\pi\hbar^2} \qquad (13.15)$$

which is the formula for the DOS of a two-dimensional electron gas. Following the same approach as in the MLG, the DOS in the presence of electron - hole puddles is found to be (Kliros 2010a, Kliros 2010b)

$$g_{BLG}(E) = \frac{2}{\pi(\hbar v_F)^2} \left[ \sqrt{\frac{2}{\pi}} s \exp\left(-\frac{E^2}{2s^2}\right) + E \, erf\left(\frac{E}{s\sqrt{2}}\right) + \frac{\gamma_1}{2} \right] \qquad (13.16)$$

Near the Dirac point, the density of states becomes

$$D_{BLG}(0) = g_s g_v \frac{m^*}{2\pi\hbar^2} \left( 1 + 2\sqrt{\frac{2}{\pi}} \frac{s}{\gamma_1} \right) \qquad (13.17)$$

which increases linearly with the potential fluctuation strength s. Consequently, the quantum capacitance of BLG can be calculated using a similar expression to Eq. (41.10).

$$C_{Q,BLG}(V_{ch}) = e^2 \int_{-\infty}^{+\infty} g_{BLG}(E) \left( -\frac{\partial f(E-E_F)}{\partial E} \right) dE \qquad (13.18)$$

Figure 13.8 shows our numerical results for the quantum capacitance of BLG for different potential fluctuation strengths at room temperature T=300 K. The numerical results from Eq. (13.18) along with the experimental data from the 'suplimentary information' in the work of (Xia et al. 2009), are shown in Fig. 13.8. In order to achieve the best fit to the available experimental data at temperature T=200 K, a standard deviation of potential fluctuations $s$=228 meV is used in combination with a 30% enhanced value of the Fermi velocity i.e., $v_F = 1.32 \times 10^6 \, m/sec$. The value of $s$ is consistent with direct calculations predicting that the fluctuation strength in BLG tends to be much larger than in MLG samples (Hwang et al. 2010). In particular, it has been estimated that $s_{BLG}/s_{MLG} \approx 32/\sqrt{\tilde{n}}$ with $\tilde{n} = n/10^{10}$, which for carrier densities $n \approx 10^{12} \, cm^{-2}$, a value of $s_{BLG} \approx 3 s_{SLG}$ is obtained. On the other hand, the Fermi velocity renormalization in bilayer graphene is proposed which corresponds to about 30% enhancement in comparison to monolayer graphene (Park et al. 2009, Chae et al. 2012). This fact is also consistent with a recent calculation in bilayer graphene that predicts a monotonic suppression of the effective mass as a function of decreasing density (Zou et al. 2011). It is clearly seen that our simple model is unable to describe the pronounced electron-hole asymmetry of the experimental capacitance curve since the induced by the electron-electron interaction difference in the effective mass of the electron and hole branches is not taken into account (Zou et al. 2011). Fimally, the temperature dependence of the quantum capacitance of BLG, adopting *s*=228 meV, is dipicted in Fig. 13.10.

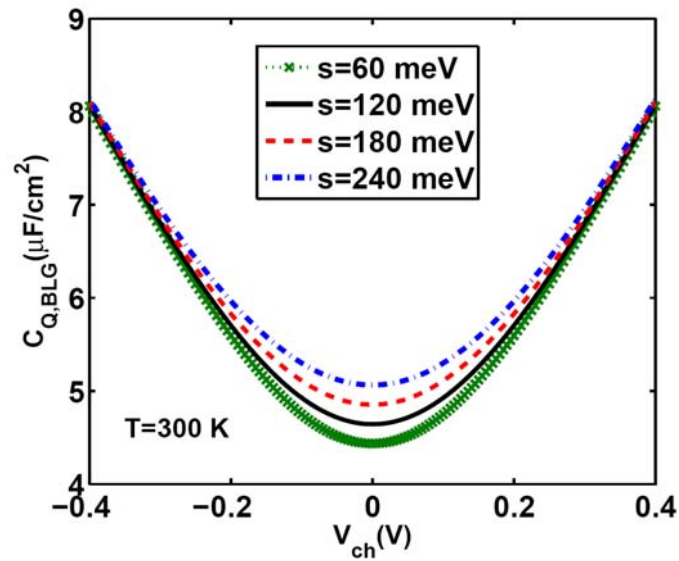

**Figure 13.8.** Calculated quantum capacitance of BLG for different potential fluctuation strengths at room temperature T=300 K.

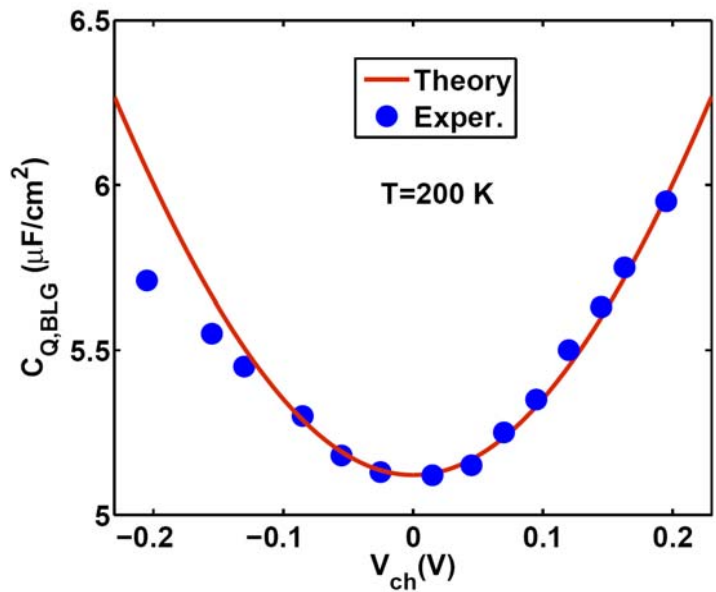

**Figure 13.9.** Fitting of the calculated quantum capacitance of BLG including electron-hole puddles to the measured one from (Xia et al., 2009).

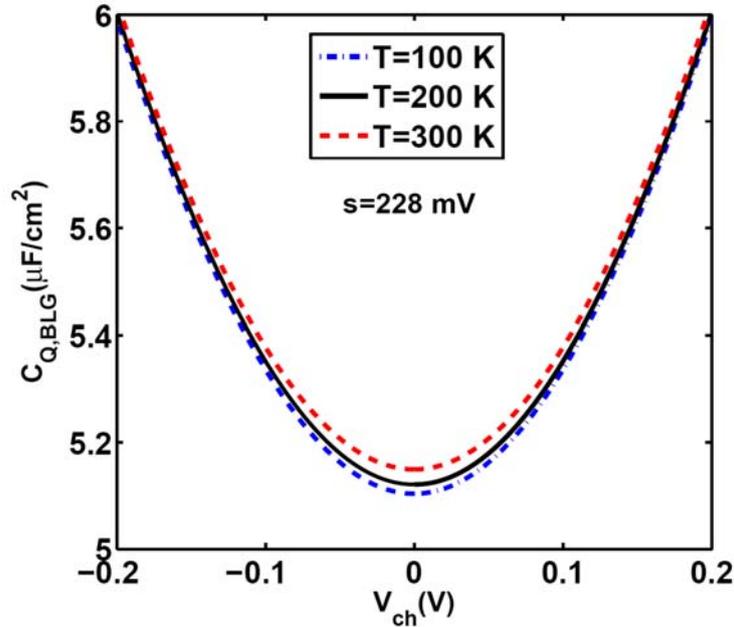

**Figure 13.10.** Temperature dependence of the quantum capacitance of BLG including electron-hole puddles. A value of $s$=228 meV is used in the calculations.

### 13.3 Quantum Capacitance of Graphene Nanoribbons

The characteristics of quantum capacitance change drastically for nanoscale ribbons cut from graphene sheets. Particular aspects of the self-consistent gate electrostatics in GNRs as well as numerical and analytical studies of quantum capacitance of GNRs have been reported in the literature (Guo et al. 2007, Shylau et al. 2009, Kliros 2010c). Recently, an interesting interplay between quantum capacitance effects in GNR and the lateral graphene side gates has been revealed, giving rise to an unconventional negative quantum capacitance (Reiter et al. 2014).
In this section, the characteristcs of the quanum capacitance of GNRs are explored with emphasis on the effect of density inhomogeneity due to the presence of electron-hole puddles (Kliros 2012). In particular, the dependence of quantum capacitance on the channel potential for different temperatures, nanorribon widths and strengths of potential fluctuation is investigated. In GNRs, the lateral confinement leads to a band gap, which furthermore is highly sensitive to the width and edge shape of the GNR (chirality). However, recent experiments suggest no clear dependence of the bandgap on the chirality due to the resolution limit of the current patterning techniques. Moreover, edge disorder effectively wipes out any distinction between zigzag and armchair GNRs (Querlioz et al. 2008).

Thus, we consider a simple physical picture where, in a GNR of width W<<L, the wave vector perpendicular to the transport direction is quantized by hard-wall boundary conditions to be $k_y = n\pi/W$ with $n = 0, \pm 1, \pm 2,...$ Then, the energy dispersion relation in the low-energy range close to the Dirac point, is given by (Fang et al. 2007)

$$E_n(k_y) = \pm \hbar v_F \sqrt{k_x^2 + \left(\frac{n\pi}{W}\right)^2} \tag{13.19}$$

indicating that both the conduction band (+) and valence band (−) split into one-dimensional subbands indexed by *n*. Note that a good agreement between the above simple picture and the tight-binding numerical data for GNRs with irregular edges has been reported (Bresciani et al 2010) validating somehow the above simple expression. The corresponding DOS of the GNR can be easily calculated according to the relation

$$D(E) = \frac{2}{\pi} \sum_n \left[\frac{dE_n(k_x)}{dk_x}\right]^{-1} \tag{13.20}$$

where the summation includes all tranverse modes and a factor 2 for the spin degeneracy is included. As a result,

$$D(E) = \frac{4}{\pi \hbar v_F} \sum_{n>0} \frac{E}{\sqrt{E^2 - E_n^2}} \Theta(E - E_n) \tag{13.21}$$

where $\Theta(E)$ is the Heaviside unit step function and $E_n$ are the subband threshold energies given by $E_n = n\pi \hbar v_F / W = n E_G / 2$ with $E_G$ the band gap of GNR. The DOS is the same for both the conduction and valence band and exhibits Van Hove type singularities at energies $E_n$ measured from the Dirac point. The quantum capacitance of the GNR can be calculated by using Eq. (41.4) and in the low temperature limit, $E_F > E_n \gg k_B T$, one obtains

$$C_{Q,GNR}(V_{ch}) = \frac{4e^2}{\pi \hbar v_F} \sum_{n>0} \frac{E_F}{\sqrt{E_F^2 - E_n^2}} \Theta(E_F - E_n) \tag{13.22}$$

where $V_{ch} = E_F / e$ is the channel potential which describes the change in the chemical potential due to the filling of the energy bands (Kliros 2010c).

It has been reported that, in GNRs, charged impurities at the graphene substrate interface create an inhomogeneous density profile breaking the system into electron-hole puddles (Adam et al. 2008). It is also numerically confirmed that, even for potential fluctuations on an atomistic scale, there is strong evidence for electron-hole puddle formation in agreement with the experimental observations (Schubert and Fehske 2012). Following the procedure of previous sections, we incorporate the presence of electron-hole puddles by expressing the DOS as convolution of the DOS of prestine GNRs with a Gaussian distribution function of standard deviation *s* (Kliros 2012), that is,

$$g_{GNR}(E) = \frac{4}{\sqrt{2\pi^3} \hbar v_F s} \sum_{n>0} \int_{E_n}^{\infty} \frac{\varepsilon}{\sqrt{\varepsilon^2 - E_n^2}} \exp\left(-\frac{(\varepsilon - E)^2}{2s^2}\right) d\varepsilon \tag{13.23}$$

Consequently, the temperature dependent quantum capacitance of the GNR, in the presence of electron-hole puddles, can be calculated by inserting Eq. (13.21) into Eq. (13.4) and performing numerically the integration.

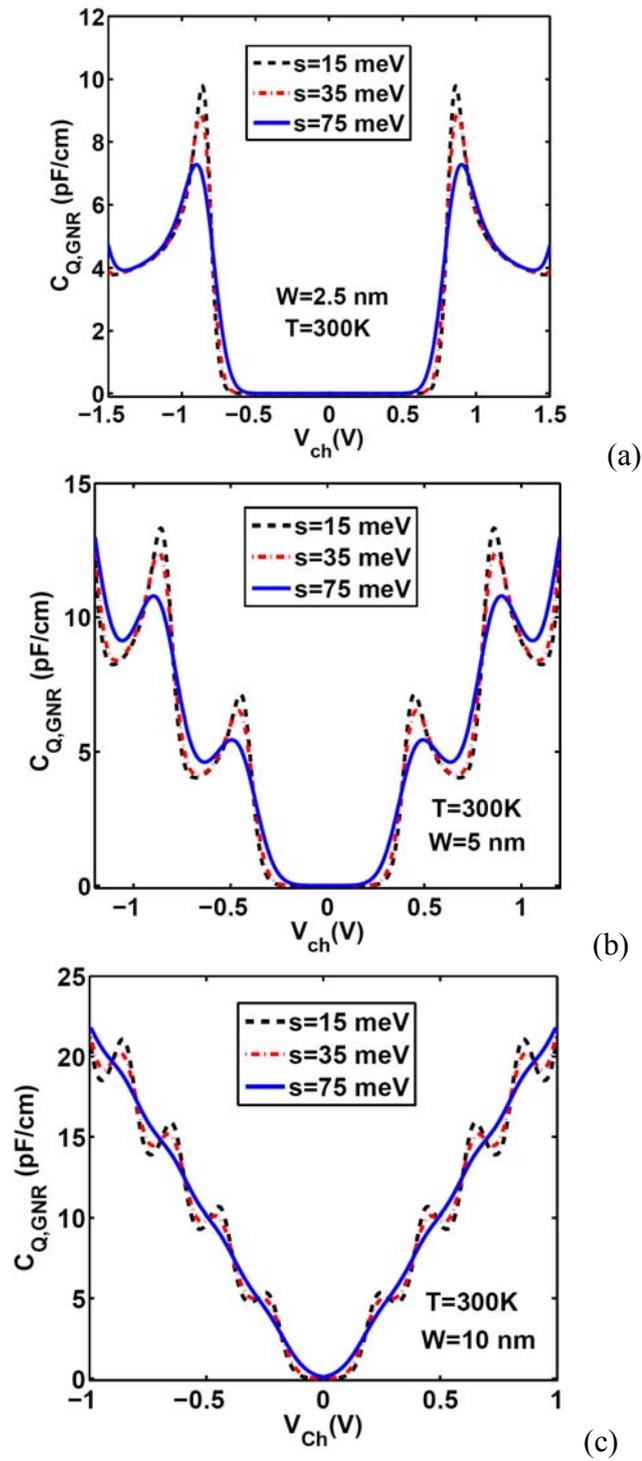

**Figure 13.11.** Calculated room temperature quantum capacitance $C_Q$ versus channel potential $V_{ch}$, for different values of fluctuation strength and nanoribbon width **a)** 2.5 nm **b)** 5 nm and **c)** 10 nm.

Fig. 13.11 (a)-(c) display the room temperature quantum capacitance as a function of the channel potential, for sub-10 nm GNRs for three different values of fluctuation strength s=15, 35 and 75 meV, which are representative values consistent with the literature (Li, Q. et al. 2011). The obtained small values of quantum capacitance are attributed to i) low DOS characterizing the atomically thin quasi-1D channel and ii) further reduction of the DOS due to quantum confinement boundary conditions in the GNR transverse direction. Because the quantum capacitance $C_Q$ is proportional to the DOS, the multipeaks in $C_Q$ are reminiscent of the Van Hove singularities and follow the population of electron subbands as the channel potential increases. As it is seen theese peaks are broadened as the fluctuation strength increases and this effect is more pronounced in wide GNRs since the distance between the peaks of the DOS decreases as the GNR's width increases. One can clearly see that the quantum oscillations are suppressed in wide GNRs of W=10 nm under potential fluctuation strength of s=75 meV. In Fig. 13.12, the room temperature quantum capacitance for GNR of different widths, is plotted as a function of channel voltage adopting a reasonable moderate value of s =35 meV. The peak value of the quantum capacitance drastically decrease as the GNR's width increases, whereas the number of peaks increases as the width increases. Moreover, the channel potential where the first peak of $C_Q$ is observed is strongly dependent on the GNR's width. Finally, Fig. 13.13 illustrates the temperature evolution of the characteristics $C_Q (V_{ch})$ for a GNR of width W=5 nm. It is observed that the peak values decrease as the temperature increases but the thermal broadening effect is small leading to nearly unchanged energy bandgap.

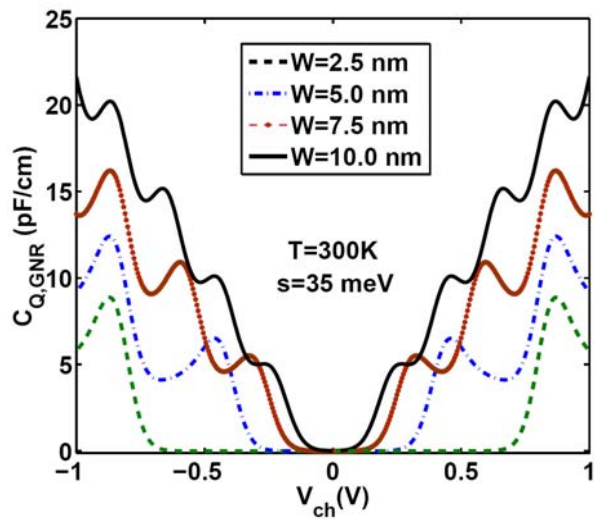

**Figure 13.12.** Room temperature quantum capacitance for GNR of different widths. A reasonable moderate value of s =35 meV is used.

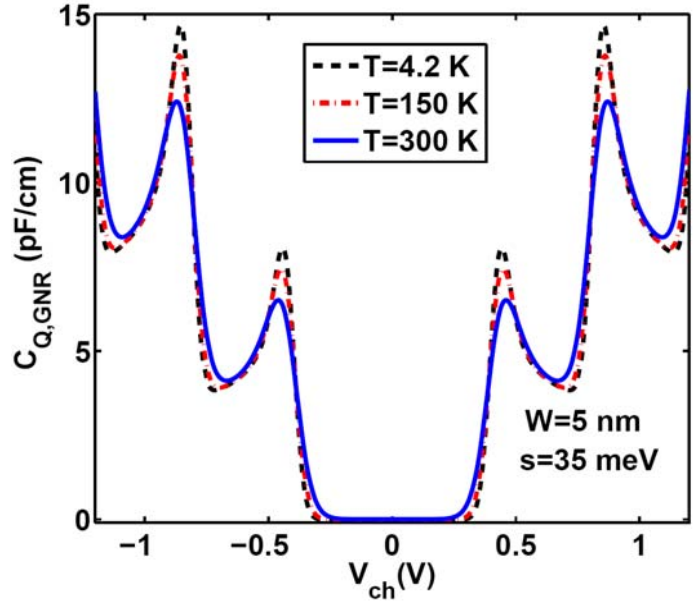

**Figure 13.13.** Temperature evolution of the quantum capacitance versus channel potential for a GNR of width W=5 nm.

## 13.4 Size and Edge - Chirality Dependent Quantum Capacitance of GNRs

In this section, an analytical model for the quantum capacitance of armchair graphene nanoribbons (AGNRs) is utilized in order to explore several size and edge effects. The model is based on the effective mass approximation which is accurate enough for narrow GNRs since their band dispersion curves are approximately parabolic. Moreover, since edge bond relaxation and third nearest neighbor (3NN) interaction have great impact on the band structure of GNRs (Gunlycke and White 2008), we have incorporated these factors in our model. It is worth noting that 2NN interaction, only shifts the dispersion relation in the energy axis but does not change the band structure and therefore, it can be ignored. Thus, the band structure of an AGNR can be expressed in the as (Gunlycke and White 2008)

$$E_n(k_x) = \pm\sqrt{E_{C,n}^2 + (\hbar v_n k_x)^2} \qquad (13.24)$$

with

$$E_{C,n} = \gamma_1\left(2\lambda\cos(n\theta)+1\right) + \gamma_3\left(2\cos(2n\theta)+1\right) + \frac{4(\gamma_3+\Delta\gamma_1)}{N+1}\sin^2(n\theta) \qquad (13.25)$$

and

$$\left(\frac{\hbar v_n}{3a_{cc}}\right)^2 = -\frac{1}{2}\lambda\gamma_1 \cos(n\theta) \times \left(\gamma_1 + \gamma_3\left(2\cos(2n\theta)+1\right) + \frac{4(\gamma_3 + \Delta\gamma_1)}{N+1}\sin^2(n\theta)\right)$$
$$-\gamma_3\left(\gamma_1 + 2\gamma_3 \cos(2n\theta) + \frac{4(\gamma_3 + \Delta\gamma_1)}{N+1}\sin^2(n\theta)\right) \quad (13.26)$$

where $\theta = \pi/(N+1)$, ± stands for the conduction and valence band respectively, N is the naumber of carbon atoms in the tranverse direction, $n$ denotes the subband index, and $E_{C,n}$ is the band edge energy of the $n$-th subband. The first set of conduction and valence bands have band index $\lambda = -1$. Moreover, the hopping parameters $\gamma_1 = -3.2$ eV, $\gamma_3 = -0.3$ eV and $\Delta\gamma_1 = -0.2$ eV refer to 1NN, 3NN and the correction to 1NN due to edge bond relaxation effect, respectively. Then, supposing electron-hole band structure symmetry, the energy gap is given by $E_{G,n} = 2E_{C,n}$ and the carrier effective mass $m_n^* = E_{C,n}/v_n^2$. It is worth noting that the effective mass has width dependence through both terms in the numerator and denominator of the last expression. The width of an AGNR can be calculated using the relation $W = (N+1)\sqrt{3}\,a_{cc}/2$ where $a_{cc}$ =0.142 nm is the graphene's lattice constant. The schematic of a typical field effect device that uses an AGNR as channel material and a thin gate insulator is dipicted in Fig. 13.14.

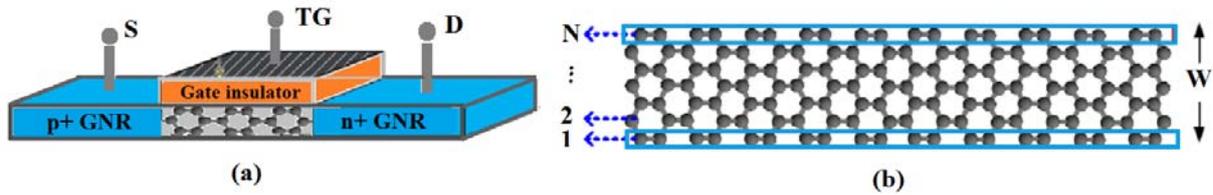

**Figure 13.14**. Schematics of a typical top-gated field effect device that uses an AGNR as channel material and a thin gate-insulator.

Based on the effective mass approximation, the density of states (DOS) of the AGNR per unit length is given by

$$D(E) = \frac{1}{\pi\hbar}\sum_{n>0}\sqrt{\frac{2m_n^*}{E-E_{C,n}}}\,\Theta\left(E-E_{C,n}\right) \quad (13.27)$$

where $\Theta(E)$ is the Heaviside unit step function. Then, the electron concentration can be evaluated as

$$n_{1D}(E_F) = \int_{E_{C,n}}^{\infty} D(E) f(E, E_F)\, dE \quad (13.28)$$

where $f(E, E_F)$ is the Fermi-Dirac distribution function. After integating Eq. (13.28) we obtain

$$n_{1D}(E_F) = \sqrt{\frac{2k_B T}{\pi \hbar^2}} \sum_{n>0} \sqrt{m_n^*} F_{-1/2}(\eta_{F,n}) \qquad (13.29)$$

where $F_{-1/2}(\eta_{F,n})$ is the Fermi integral of order (-1/2) and $\eta_{F,n} = (E_F - E_{C,n})/k_B T$. Then, the quantum capacitance of AGNR can be calculated as $C_Q = e^2 \, \partial n_{1D}/\partial E_F$. Finally, using Eq.(13.29) and writing in terms of Fermi integrals of order (-3/2), we obtain the quantum capacitance per unit length (Kliros, 2013)

$$C_Q = e^2 \sqrt{\frac{2}{\pi \hbar^2 k_B T}} \sum_{n>0} \sqrt{m_n^*} F_{-3/2}(\eta_{F,n}) \qquad (13.30)$$

where the Fermi integral of order j is defined by

$$F_j(\eta) = \frac{1}{\Gamma(j+1)} \int_0^\infty \frac{t^j}{1+\exp(t-\eta)} dt \qquad (13.31)$$

and $\Gamma(x)$ represents the Gamma function.

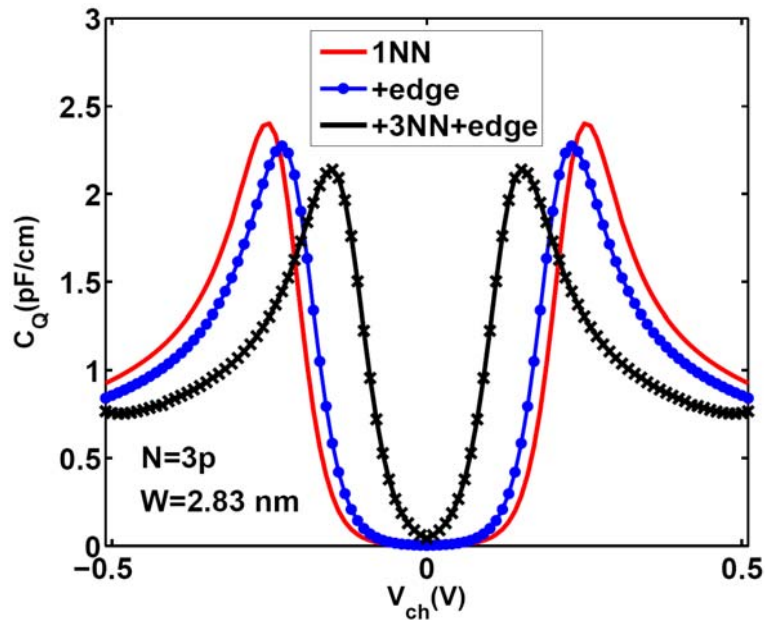

(a)

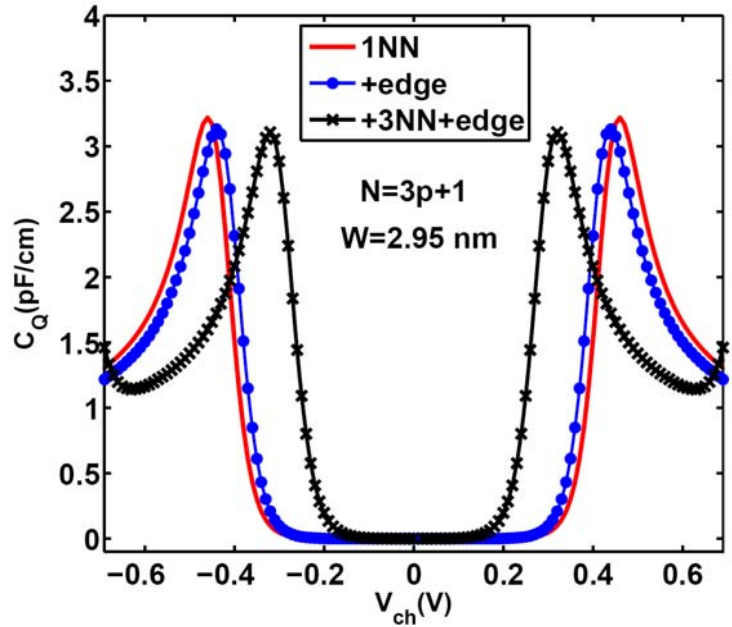

**Figure 13.15.** Characteristics $C_Q(V_{ch})$ for (**a**) an 25-AGNR (N=3p) and (**b**) an 26-ARGN (N=3p+1).

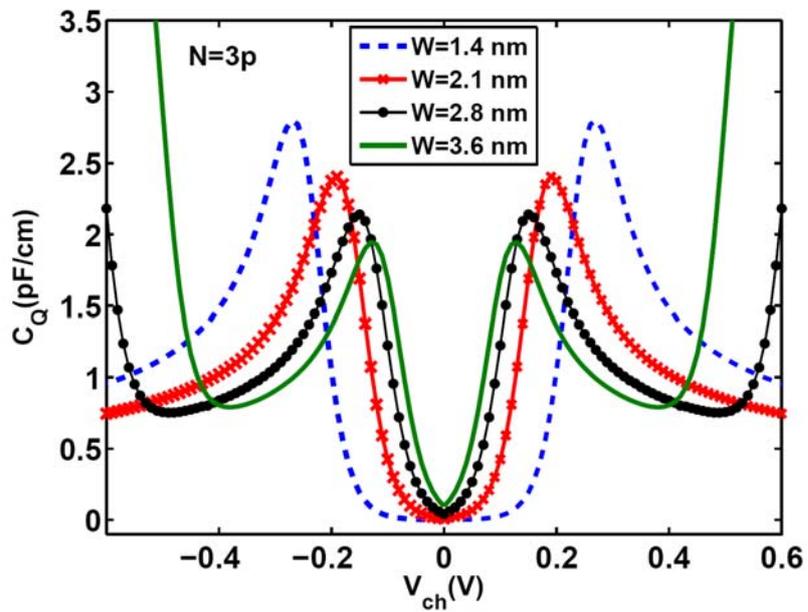

(a)

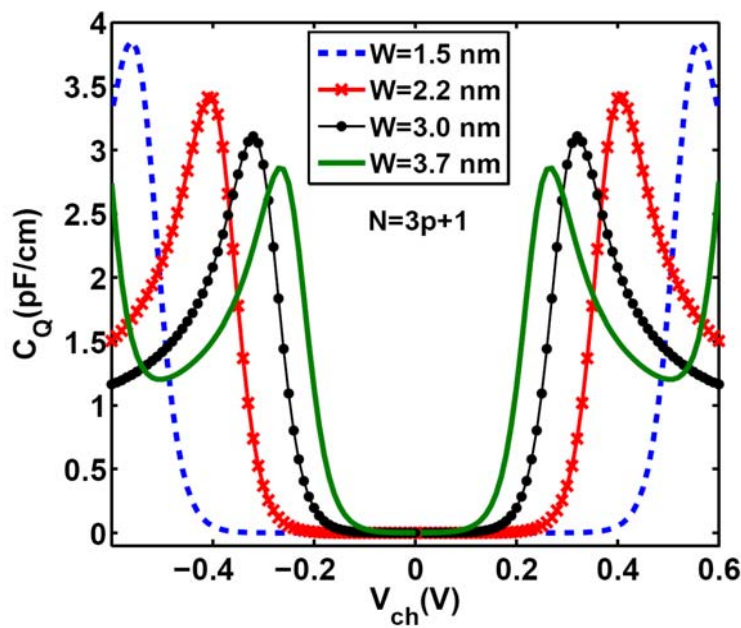

(b)

**Figure 13.16.** Characteristics $C_Q(V_{ch})$ of GNR with increasing widths for the two families **a)** N=3p and **b)** N=3p+1.

We proceed now with the investigation of the effect of 3NN interactions and edge bond relaxation correction on the quantum capacitance versus channel potential $V_{ch} = E_F/e$ using the result of Eq. (13.30). It has been demonstrated that in the presence of both 3NN interactions and edge bond relaxation correction, all AGNRs are semiconducting with bandgaps well separated in to three different groups N = 3p, N = 3p + 1, N = 3p + 2 (Wang, 2008). However, the bandgap of the family N = 3p + 2 is significantly reduced resulting in a close-to-metallic channel. Therefore, we restrict our study in the groups N = 3p and N = 3p+1 that are more promising for nanoelectronic applications having larger bandgaps. Figures 13.15 (a) and (b) display the characteristics $C_Q(V_{ch})$ for an 25-AGNR (N=3p) and an 26-ARGN (N=3p+1) respectively. The family N=3p+1 shows larger peak values of the quantum capacitance and larger band gaps than the family N=3p. In both families, the edge bond relaxation correction has small effect on the bandgap whereas, the 3NNN-interaction drastically reduses it. Both edge bond relaxation and 3NNN reduce the peak values of $C_Q$ in the family N=3p but leave the peaks almost the same in the family N=3p+1. Figures 13.16(a) and (b) show the characteristics $C_Q(V_{ch})$ of GNR with increasing widths for the two families N=3p and N=3p+1, respectively. Moreover, the peak values of $C_Q$ drastically decrease as the GNR's width is increased and the value of $V_{ch}$ in which each peak appears, is stronly dependent on the GNR's family. Finally, Fig. 13.17 compares the dependence of $V_{ch}$ at the first quantum capacitance peak on the GNR's width for the two families N=3p and N=3p+1 whereas the same comparison is illustrated in Fig. 13.18 for the width dependent quantum capacitance peaks. As it is seen, the peak values of quantum capacitance are observed at channel potentials that are inversely proportional to the widths and the same relation is obeyed by the peak values of quantum capacitance versus GNR's width for both GNR families. However, for a given width, family N=3p+1 presents larger peak values of quantum capacitance that also correspond to larger values of channel potential compared to family N=3p.

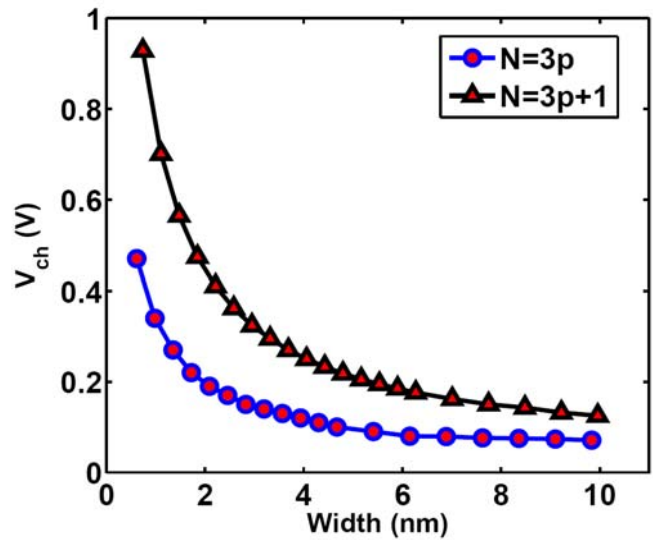

**Figure 13.17.** Dependence of $V_{ch}$ at the first quantum capacitance peak on the GNR's width for the two families N=3p and N=3p+1.

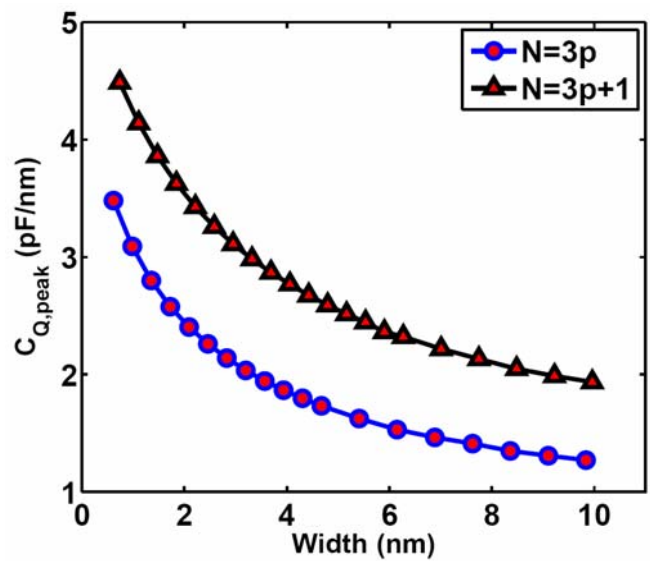

**Figure 13.18.** Dependence of the first quantum capacitance peak on the GNR's width for the two families N=3p and N=3p+1.

## 13.5 Concluding Remarks

Quantum capacitance - voltage characteristics are important for understanding the device physics and assessing the performance of graphene based nanodevices when operate in the quantum capacitance limit where the quantum capacitance dominates the total gate capacitance. On the other hand, the quantum capacitance effect can be utilized to design high quality factor varactors and other types of high performance sensors. Moreover, if the gate-tunable quantum capacitance is coulpled with a quantum inductance, the resultant circuit can be utilized as all-electronic ultra high-frequency oscillator. As a consequence, we have been motivated to present in this chapter semi-analytical models for the calculation of the quantum capacitance of both monolayer and bilayer graphene and its nanoribbons. Since electron-hole puddles are experimental facts in all graphene samples, they have been incorporated in our calculations. The temperature dependence of the quantum capacitance around the charge neutrality point is also investigated and the obtained results are in agreement with many features recently observed in quantum capacitance measurements on both monolayer and bilayer graphene devices. Furtheremore, the impact of finite-size and edge effects on the capacitance of graphene nanoribbons is studied taking also into account both the edge bond relaxation and third-nearest-neighbour interaction in the band structure of GNRs.

In conclusion, the charge density inhomogeneity due to the presence of electron-hole puddles affects significantly the values of the quantum capacitance of both monolayer and bilayer graphene devices around the Dirac point as well as the peak values of the oscillating quantum capacitance of graphene nanoribbons versus the channel potential. Therefore, charge fluctuations should be taken into account when studing the electronic properties of graphene nanodevices at low densities. On the other hand, in order to achieve the best fit to the measured quantum capacitance of bilayer graphene a renormalization of Fermi velocity is adopted in accordance to recent theoretical predictions. Concerning the non-monotonic behaviour of the quantum capacitance of sub-10 nm GNRs versus channel potential, its peak values and corresponding voltages are strongly dependent on the ribbon's width and are distinct for the two semiconducting AGNR families $N=3p$ and $N=3p+1$. Such dependence could be utilized to determine the GNR's width from quantum capacitance measurements.